\documentclass[twocolumn,showpacs,prb,amsmath,amssymb]{revtex4}

\def\fakebold#1{\relax\ifvmode\leavevmode\fi%
\ifmmode%
\setbox0=\hbox{$#1$}%
\else%
\setbox0=\hbox{#1}%
\fi%
\kern-.02em\copy0 \kern-\wd0%
\kern .04em\copy0 \kern-\wd0%
\kern-.0125em\raise.02em\box0%
}%

\newcommand{\bfpi}{\mbox{\fakebold{$\pi$}}}
\newcommand{\bfpartial}{\mbox{\fakebold{$\partial$}}}

\usepackage{graphicx}
\usepackage{bm}
\begin{document}

\title{Renormalization of the upper critical field by superconducting fluctuations}

\author{V. M. Galitski}
\author{S. \surname{Das Sarma}}
\affiliation{Condensed Matter Theory Center and Center for
Superconductivity Research, 
Department of Physics, University of Maryland, College Park, Maryland
20742-4111}

\begin{abstract}
We study the effect of superconducting fluctuations on the upper critical field of a disordered superconducting
film at low temperatures.  The first order  fluctuation correction is found explicitly.
In the framework of the perturbative analysis, superconducting fluctuations are shown to shift
the upper critical field line toward lower fields and do not lead to an upward curvature.
Higher order corrections to the quadratic term coefficient in the Ginzburg-Landau free energy
functional are studied. We extract a family of the mostly divergent diagrams and formulate
a general rule of calculating a diagram of an arbitrary order. We find that the singularity  gets more
severe with increasing perturbation theory order.
We conclude that the renormalization of the Ginzburg-Landau coefficients  by superconducting  fluctuations 
is an essentially non-perturbative effect. As a result, the genuine transition line  may be strongly shifted from the
classical mean-field curve in a two-dimensional superconductor.
\end{abstract}


\maketitle

\section{Introduction}

The magnetic field-temperature phase diagram of type II
superconductors has been of experimental and theoretical interest for a
very long time.  Much interest has recently focused on the disordered
two-dimensional superconductors partly motivated by the two-dimensional
nature of the high-$T_c$ cuprates although disordered thin films of
``ordinary'' superconductors have been studied for years.  The subject
matter of this paper is the temperature dependence of the upper critical
field, $H_{c2}(T)$, above which the system goes normal.  In particular, we
consider the problem of superconducting fluctuation effects on the upper
critical field of a disordered superconducting film at low
temperatures.

There have been a number of experiments studying the behavior of the upper critical
field line as a function of temperature and disorder.
Some of them\cite{exp} have shown strong deviations from the classical mean-field theory, \cite{HW}
such as an upward curvature in the $H_{c2}(T)$-line.
These experiments have induced a considerable theoretical interest in the subject.

Let us start with recalling the previous theoretical studies. The first attempt to
explain the observed effects has been  made by Golubov and Dorin \cite{GD} who considered the combined effects
of disorder and Coulomb interaction in first-order perturbation theory.  Golubov and Dorin's calculations predicted some upward curvature 
in the upper critical field line. However, later re-examination of the Golubov and Dorin's
theory showed some very technical deficiencies in the original calculations.
Namely, Smith {\em et al.}  \cite{Smith} have performed both first-order perturbation theory calculations
and Oreg and Finkelshtein renormalization group treatment \cite{OF} of the problem. Their
careful analysis has not shown any anomalous upward curvature in either case.
So, the final verdict was that disorder and Coulomb interaction together could
not explain the observed deviations from the mean field results.

Zhou and Spivak have suggested \cite{ZS} a mechanism of  $H_{c2}$ enhancement based on a
possibility of the spontaneous formation of a granular-like superconducting structure 
in a disordered film at a low enough temperature. The  idea is that optimal
fluctuations in the distribution of impurities  may lead to the formation of local
regions where the local critical field exceeds the system-wide average value. 
At a low enough temperature, these disorder-induced superconducting droplets
form a Josephson network allowing for the global superconductivity in the film. 
Recently, Galitski and Larkin \cite{GaL} have included the effects of quantum fluctuations
into the picture and showed that the ``mesoscopic disorder'' 
({\em i.e.}, usual non-magnetic impurities) was very weak to produce a considerable effect on 
the critical field line and the global superconducting glassy state suggested 
by Zhou and Spivak \cite{ZS} should be destroyed by quantum fluctuations. However,
the ``pinning disorder'' (grain boundaries, dislocation clusters, {\em etc.})
may be strong enough to compete with the quantum fluctuations. 
The prediction is that the anomalous upward curvature due to the optimal
disorder fluctuations is possible but the corresponding effect is 
non-universal and depends on the pinning properties of the system.
Let us mention a Comment \cite{Ikeda_Com} on the paper \cite{GaL} by Ikeda,
who emphasized a possible importance of superconducting fluctuations. 
The purpose of the present work, which is partially motivated by Ref.[\onlinecite{Ikeda_Com}], 
is to explicitly evaluate fluctuation corrections  to the upper critical field line at low temperatures. 

Let us  emphasize that the definition of the superconducting transition point itself  is
a very delicate issue in a two-dimensional system. Strictly speaking, at finite fields,
superconductivity as a state with zero resistivity is never achieved which
is basically a manifestation of the general Mermin-Wagner theorem.\cite{MW}
In real experiments, what is observed is a drop of the resistance
to some very small but finite values. Such a point is usually
defined to be the transition point. Further decrease of the  temperature
or external field yields a very slow decay of the resisitivity.
From the theoretical point of view, the definition of the critical point is also non-trivial.
One can define the transition point as a point at which the superfluid 
density $\rho_{\rm s}$ becomes non-zero. Another approach is  to study
the behavior of the quantity $\left\langle \psi(0) \psi({\bf r}) \right\rangle = \Delta({\bf r})$ by the use
of Ginzburg-Landau type expansions and define the transition point
as a point at which  the quadratic term
vanishes. There are no solid grounds to believe that these 
two approaches are equivalent. In the present paper, we follow the
latter approach, which, as we think, should  correspond to the
point at which the critical drop in the resistance to some finite value  takes place. 

The question of  $T_c$ renormalization in zero field has been
recently addressed by Larkin and Varlamov in the Review [\onlinecite{LV}]. 
The two-dimensional result for the shift of the transition temperature was found as 
$\delta T_c / T_c  \sim \left[ {\rm Gi \,} \ln{\rm Gi} \right]$, where ${\rm Gi}$ is the Ginzburg
parameter, which in a disordered two-dimensional superconductor
is of order of the inverse conductance. Thus, one concludes
that the effect of fluctuations in zero field is quite small and the 
fluctuations can move the transition point only within an extremely
narrow Ginzburg region. We will return to the discussion of  the
validity of this conclusion at the end of the paper.

Our paper is structured as follows: 
In Sec. II, we formulate the Ginzburg-Landau theory for the case
of a strong external magnetic field. We emphasize that the Ginzburg-Landau
expansion in this case can be made with respect to the modulus of the order parameter only and
not on its spatial gradients. However, one can still write down
a formal operator expansion. The vanishing of the lowest Landau 
level matrix element of the operator $\hat A$ 
appearing in the quadratic term, corresponds to the transition point.
In Sec. III, we study the effect of superconducting fluctuations 
on the upper critical field in first order perturbation theory.
We calculate the first correction explicitly. We find that the first correction alone can not lead to a monotonous
upward curvature of the $H_{c2}(T)$ line but can yield a decrease of the
critical field compared to the mean-field value.\cite{HW} In Sec. IV, we study
the higher order perturbation theory corrections.
We extract a family of the mostly divergent diagrams the effect
of which parametrically exceeds the first and second order contributions. The first term
in the family appears in third order perturbation theory only. We show
that the singularity gets progressively stronger in higher orders. 
We conclude that the renormalization of the Ginzburg-Landau coefficients and, consequently, the transition line by
superconducting fluctuations is an essentially  non-perturbative effect. 
 
\section{Ginzburg-Landau expansion in strong fields}

\begin{figure*}
\includegraphics[width=3in]{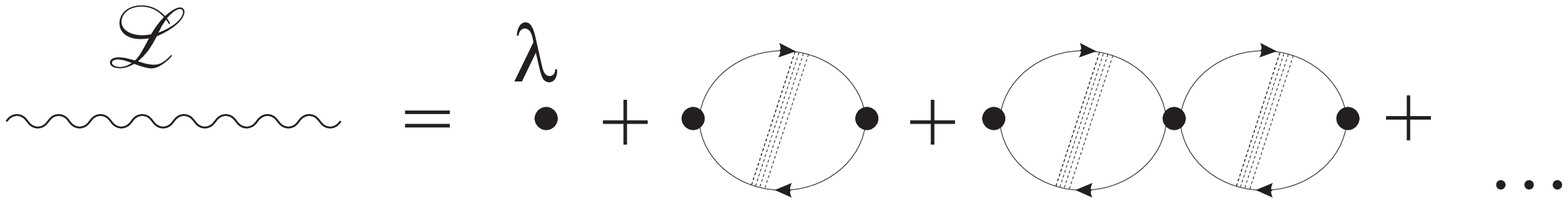}
\caption{\label{fig:fig1} Diagrammatic equation for the fluctuation propagator (\protect\ref{L}) (curly line).}
\end{figure*}  

\begin{figure*}
\includegraphics[width=3in]{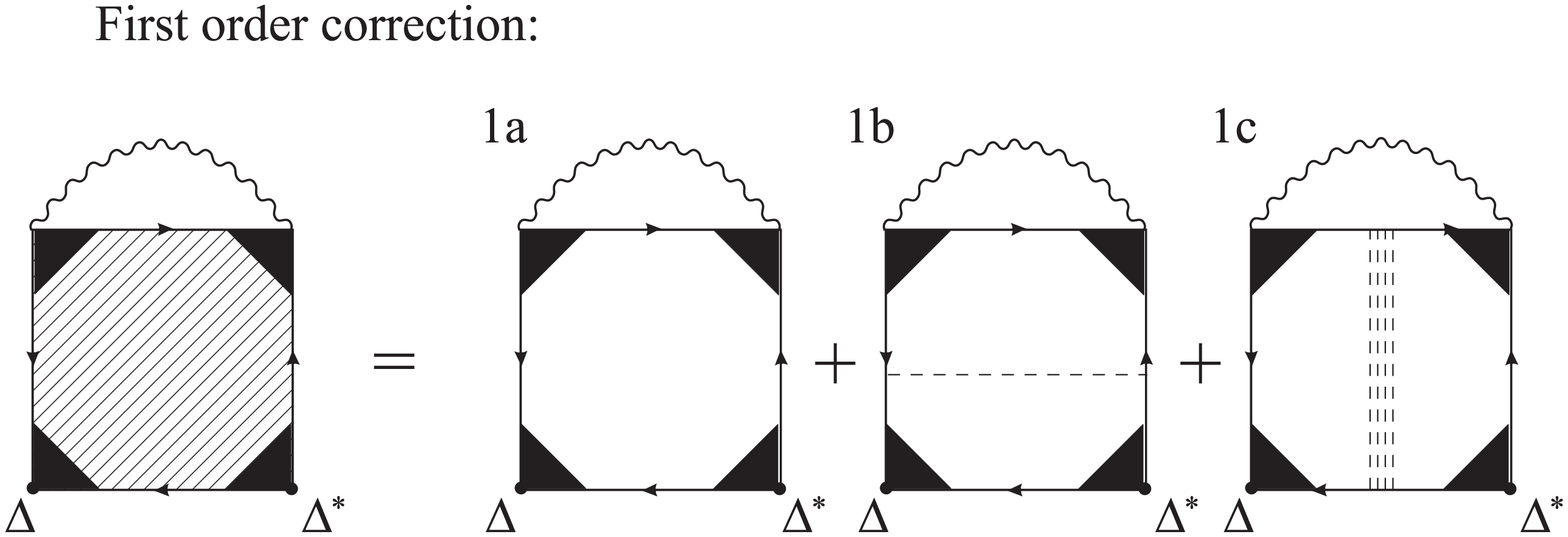}
\caption{\label{fig:fig2} First order fluctuation corrections to the quadratic part of the Ginzburg-Landau free energy expansion.}
\end{figure*}  

It is well-known that a superconductor in a relatively weak magnetic field can be described
with the aid of the Ginzburg-Landau theory. The Ginzburg-Landau expansion of the free energy 
has the following standard form:
\begin{eqnarray}
\label{GL1}
{\cal F}\left[ \Delta({\bf r}) \right] = {\cal F}_{\rm N} +
\int {d^2 {\bf r}}\, \Biggl\{ \!\!\!\!\!\!\! && a \left| \Delta({\bf r}) \right|^2 
+ b \left| \Delta({\bf r}) \right|^4 \\   
&&
+ {1 \over 4m} \left| \left[ \nabla - 2 i e {\bf A}({\bf r}) \right] 
 \Delta({\bf r}) \right|^2  \Biggr\},
\nonumber 
\end{eqnarray}
where ${\cal F}_{\rm N}$ is the free energy of a normal metal, $\Delta({\bf r})$ is the superconducting 
order parameter, $a$, $b$, and $m$ are some coefficients which  depend both on temperature and magnetic field.
Coefficient $a$ vanishes at the transition point.

To derive Eq.(\ref{GL1}) from the microscopic theory, {\em i.e.} from the initial BCS hamiltonian,
one has to fulfill the following standard steps:
First, the quartic interaction term should be decoupled with a Hubbard-Stratonovich field $\Delta$.
Then, the functional integral on the electron degrees of freedom becomes Gaussian and can be easily evaluated.
The resulting action $S(\Delta)$ is quite complicated, however in the vicinity of the superconducting
transition, $\Delta$ is small and the action can be expanded on it. If one is interested in the equilibrium
thermodynamic properties of a superconductor, the time-dependence of $\Delta$ may be suppressed
and one can talk of deriving an effective free energy functional rather than of an effective action.
However, another essential assumption has to be made in order to get Eq.(\ref{GL1}).
Namely, one has to suppose that the spatial variations of the order parameter are small enough 
so that term $\left|\left( \nabla - 2 i e {\bf A}({\bf r}) \right) 
 \Delta({\bf r}) \right|$ is  small in some sense. Finally, expanding both on the order parameter
 and on its spatial gradients on can derive the Ginzburg-Landau free energy functional (\ref{GL1}) the minimum
 of which yields the famous Ginzburg-Landau equations.

The above arguing partially fails in strong fields when the spatial dependence of the order parameter
becomes important. In strong fields, the expansion on $\left|\left( \nabla - 2 i e {\bf A}({\bf r}) \right) 
 \Delta({\bf r}) \right|$ is no longer possible. However, one can still expand on the magnitude of the order parameter.
 The corresponding saddle point equation can be written as:
 \begin{eqnarray}
 \label{GL2}
 {1 \over \lambda} \Delta({\bf r}) = && 
 \!\!\!\!\!\!\!
 \int C({\bf r},{\bf r}') \Delta({\bf r'}) d^2{\bf r}' \\
&& 
 \nonumber
\!\!\!\!\!\!\!\!\!\!\!\!\!\!
- \int B({\bf r},{\bf r}_2,{\bf r}_3,{\bf r}_4)  \Delta^*({\bf r}_2) \Delta({\bf r}_3) \Delta^*({\bf r}_4) 
  d^2{\bf r}_2 d^2{\bf r}_3 d^2{\bf r}_4. 
 \end{eqnarray}
Here $\hat{C}$ is Cooperon (see Fig.~1) which is a linear operator in the case under consideration.
In the presence of a magnetic field the operator can be written as
$$
 C({\bf r},{\bf r}') = T \sum_\varepsilon {\cal G}_\varepsilon({\bf r}) 
{\cal G}_{-\varepsilon}({\bf r}'),
$$ 
where ${\cal G}$ is the electron Green function in the magnetic field and
$\varepsilon$  is the fermion Matsubara frequency.

Non-linear operator $B$ is described by the square diagrams similar to the ones shown in Fig.~2.
Let us note, that this operator is not singular at the transition point and 
in its vicinity and can be considered as a local quantity for a dirty
superconductor [see explicit expression (\ref{HB}) below].


In the semiclassical approximation, the magnetic field dependence in the Cooperon can be factored
out:
\begin{equation}
\label{expA}
 C({\bf r},{\bf r}') =  C^{(0)}({\bf r} - {\bf r}') \exp{
 \left\{ - 2 i e \int\limits_{\bf r}^{{\bf r}'} {\bf A} ({\bf s}) d{\bf s} \right\}
 },
\end{equation}
where $C^{(0)}({\bf r})$ is the Cooperon without magnetic field.
Formally, operator $\hat C$ can be written in the following form:
\begin{equation}
\label{Cform}
\hat C = \int C^{(0)}({\bf r}) \exp{\left\{ - i {\bf r} {\hat \bfpi} \right\}} d^2{\bf r},
\end{equation}
where ${\hat \bfpi} = -i {\bf \nabla} - 2 e {\bf A}(\hat {\bf r})$ is the operator of the kinetic
momentum, which can be expressed in terms of the creation and annihilation operators in the Landau
basis.
Again, in small fields, one can expand the exponent in Eq.~(\ref{Cform}) which will result in 
the gradient term ${\hat \bfpi}^2$ [see Eq.({\ref{GL1})]. In strong fields, we can not do such
an expansion.

The transition point is defined as a point at which the coefficient in front of the $\Delta^2$ term 
vanishes. The corresponding operator has the form:
\begin{equation}
\label{A}
\hat A = \lambda^{-1} - \hat C.
\end{equation}
The inverse operator can be easily recognized as the pairing vertex which corresponds
to the ladder summation as shown in Fig.~1. 
\begin{equation}
\label{L}
\hat {\cal L}(0) = \hat A^{-1} = \lambda \sum\limits_{n=0}^{\infty} \lambda^n {\hat C}^n.
\end{equation}
The divergence of this quantity at the transition point corresponds to the BCS instability.

It is possible to calculate pairing vertex ${\hat {\cal L}}(\Omega)$, which is also called the fluctuation propagator,
as a function of the total energy in 
the Cooper channel $\Omega$ at arbitrary magnetic fields. From Eq.~(\ref{Cform}), one can see that
the Cooperon can be written as a series containing even powers of kinetic momentum ${\hat \bfpi}$ only. 
Thus, the Cooperon is a diagonal operator in the Landau basis, which makes [see Eqs.~(\ref{A},\ref{L})] 
the  fluctuation propagator diagonal as well. The corresponding matrix elements have the following explicit form:
\begin{eqnarray}
\label{Ln}
 \nonumber
{\cal L}_n(\Omega) = {1 \over N(0)}
\Biggl\{ &&  \!\!\!\!\!\!\! \ln{T \over T_{c0}} + \psi \left[{1 \over 2} + 
{\Omega_H \left(n + {1 \over 2} \right) + \left| \Omega \right| \over 4 \pi T} \right] \\ 
&&  \!\!\!\!\!\!\! - \psi \left[ {1 \over 2} \right] \Biggr\}^{-1},
\end{eqnarray}
where $N(0)$ is the density of states at the Fermi-line, $T_{c0}$ is the transition temperature
in zero field, $\Omega = 2 \pi n T$ is the Matsubara frequency, and 
$\Omega_H = 4 e {\cal D} H$, with ${\cal D}$ being the diffusion coefficient and
$H$ the external magnetic field. The upper critical field line is defined by the condition:
\begin{equation}
\label{MFeq}
A_0 = {\cal L}_0^{-1}(0) = 0,
\end{equation}
where index ``0'' corresponds the lowest Landau level matrix element.
The above equation yields the well-known Gor'kov's mean-field curve for $H_{c2}(T)$
(see solid line in Fig.~3).

\begin{figure*}
\includegraphics[width=3in]{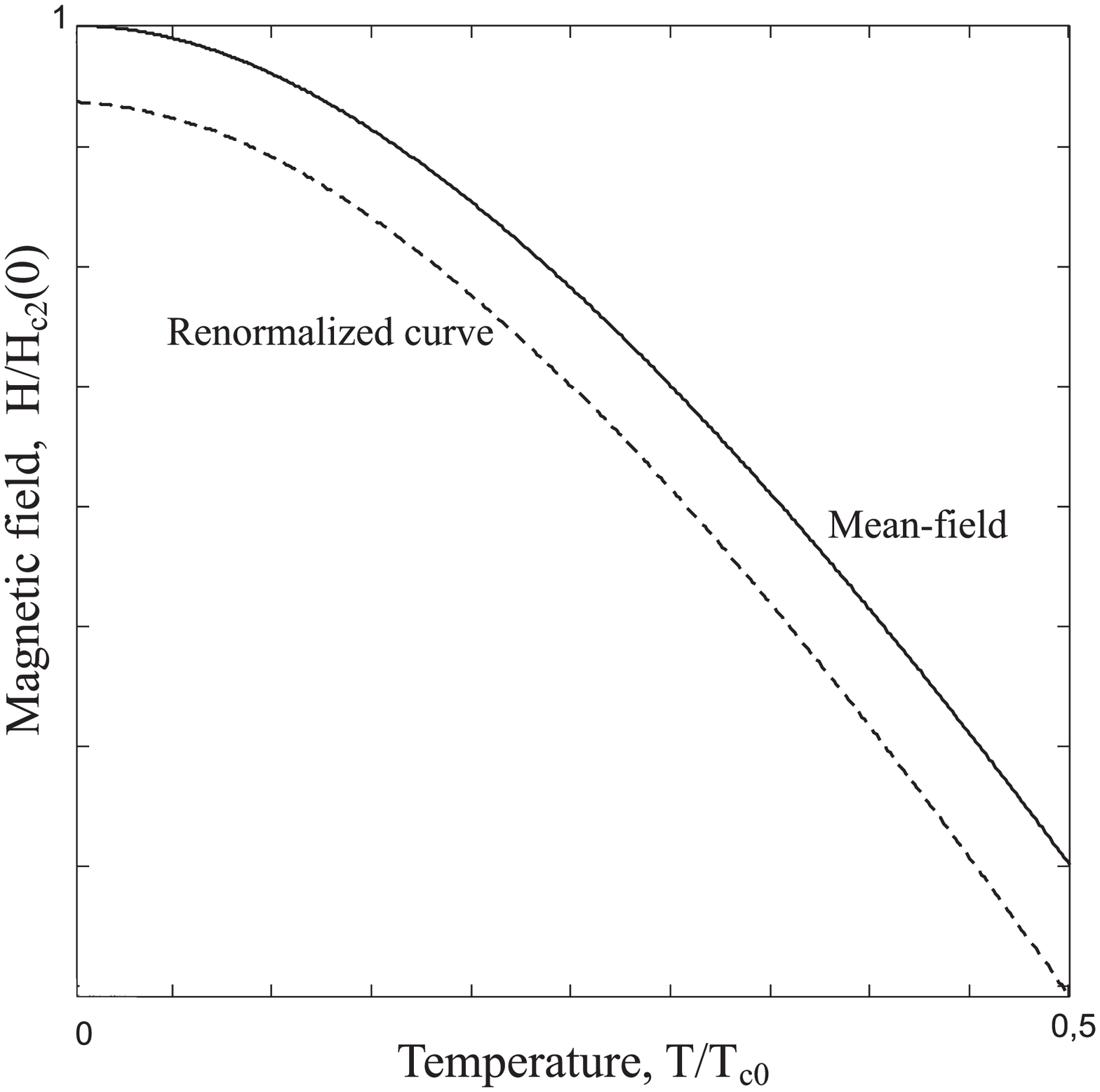}
\caption{\label{fig:fig3} Renormalization of the upper critical field by superconducting fluctuations.
The solid line shows the mean-field Gor'kov-Helfand-Werthamer result \protect\cite{HW}. The dashed line
is the upper critical field with the account for the superconducting fluctuations in first order perturbation theory.}
\end{figure*}

In order to study the effect of  superconducting fluctuations 
on the $H_{c2}$ transition line, one has to go beyond the mean field 
theory. The first fluctuation correction is described by the diagrams 
shown in Fig.~2. The corresponding correction to the Ginzburg-Landau
coefficient is
\begin{equation}
\label{da}
\delta A^{(1)} ({\bf r}_1,{\bf r}_2) = 
T^2 \sum\limits_{\Omega, \varepsilon} B_{\varepsilon,\,\Omega}({\bf r}_1,{\bf r}_2,{\bf r}_3,{\bf r}_4)
{{\cal L}}_{\Omega} ({\bf r}_3,{\bf r}_4),
\end{equation}
where $B$ is the impurity box (see Fig.~2) and ${\hat{\cal L}}_{\Omega}$ is the fluctuation propagator
introduced in the previous section.

Let us note that operator $\delta {\hat A}$ in Eq. (\ref{da}) is diagonal in the Landau representation, since it can be written in 
the form analogous to Eq.(\ref{expA}). The fluctuation propagator can be expanded on the Landau
wave-functions as follows (we use the Landau gauge here):
\begin{equation}
\label{Lexpan}
{\cal L} ({\bf r}_1,{\bf r}_2) = \sum_n {\cal L}_n \int {dp \over 2 \pi} \psi_{np}({\bf r}_1) \psi_{np}({\bf r}_2),
\end{equation}
where 
\begin{equation}
\label{psi}
\psi_{np}({\bf r}) = {\rm e}^{i p y}\, \phi_n^{\rm (osc)}(x - x_0),\,\,\, x_0 = p L_H^2, 
\end{equation}
with $\phi_n^{\rm (osc)}(x)$ being the harmonic oscillator eigenfucntion
\begin{equation}
\label{phi}
\phi_n^{\rm (osc)}(x) = {1 \over \sqrt{2^n n!}}\, \left[ {1 \over \pi L_H^2} \right]^{1/4}
{\rm e}^{-{x^2 \over 2 L_H^2}}\, {\rm H}_n \left({x \over L_H}\right), 
\end{equation}
and $L_H = \left( 2 e H \right)^{-1}$ the magnetic length for the fluctuating Cooper pairs.

Thus, the first fluctuation correction can be written as:
\begin{eqnarray}
\label{da0}
\delta A_0^{(1)} = T^2 \sum\limits_{\varepsilon, \Omega} \sum\limits_n &&  \!\!\!\!\! {\cal L}_n(\Omega)
  \int \prod\limits_{i=1}^4 d^2{\bf r}_i\,\, \nonumber
B_{\varepsilon,\,\Omega}({\bf r}_1,{\bf r}_2,{\bf r}_3,{\bf r}_4) \\
&& \,\,\,\,\, \times \nonumber
\int {dk\, dk' \over \left( 2 \pi \right)^2}\, \psi_{n\, k}\left( {\bf r}_1 \right)
 \psi^*_{n\, k}\left( {\bf r}_2 \right) \\
 &&  \,\,\,\,\,\,\,\, \,\,\,\,\,\,\,\, \times \,\, 
 \psi_{0\, k'}\left( {\bf r}_3 \right) \psi^*_{0\, p}\left( {\bf r}_4 \right).
\end{eqnarray}

At the superconducting transition point, only the lowest Landau level matrix element
of the fluctuation propagator is  divergent. In the vicinity of the transition it has the
simple form:
\begin{equation}
\label{L0}
{\cal L}_0(\Omega) = {1 \over N(0)}\, \left\{ {\left[ H - H_{c2}(T)\right] \over H_{c2}(0)} + 
{2 \left| \Omega \right| \over \Omega_H} \right\}^{-1},
\end{equation}
where $H_{c2}(T)$ is the mean-field value of the upper critical field.
Since only the lowest Landau level provides a divergent contribution, one can keep
just one term in the sum over $n$ in Eq.(\ref{da0}). With the same accuracy, one can put
$\Omega = 0$ in the expression for the box diagrams as the divergence comes from the small
values of $\Omega$ in the sum. In this approximation, the ladder summation in diagram ``c'' in Fig.~1 is not
necessary and we are left with the Hikami-box like diagram, the result for which is known:\cite{deG}
\begin{eqnarray}
\label{HB}
&&
B_{\varepsilon\, \Omega=0} ({\bf r}_1,{\bf r}_2;{\bf  r}_3,{\bf r}_4) =
{\pi N(0) \over 2} \left\{ \prod\limits_{k=1}^4
{1 \over |\varepsilon| +{1 \over 2} {\cal D} \bfpartial_{(k)}^2
}\right\} \nonumber \\
&& \times  \delta({\bf r}_1 - {\bf r}_2) \delta({\bf r}_1 - {\bf
r}_3) \delta({\bf r}_1 - {\bf r}_4) \nonumber
\\
&& \times  \Biggl[ \Biggr. |\varepsilon|  + {1 \over 8} {\cal D}
\Bigl( \Bigr. \left[ \bfpartial_{(1)} - \bfpartial_{(3)} \right]^2 +
\left[ \bfpartial_{(2)} - \bfpartial_{(4)} \right]^2 \Bigl. \Bigr)
\Biggl. \Biggr],  
\end{eqnarray}
where 
$$
\bfpartial_{(k)} = -i {\nabla} - 2 e (-1)^k {\bf A}({\bf r}).
$$
In Eq.(\ref{HB}), the factors $\left[|\varepsilon| +{1 \over 2} {\cal D} \bfpartial^2 \right]^{-1}$ come
from the four impurity vertices, three delta-functions correspond to the locality of the quantity in the
dirty limit ($\tau T_{c0} \ll 1$, where $\tau$ is the scattering time), and the last factor comes from
calculating the square diagrams themselves. Let us note that the leading frequency and momentum
independent terms are cancelled out and one has to keep $\epsilon $ and ${\bf q}_i$ finite to get a non-zero
contribution. This property can be proven not only for box-like diagrams shown in Fig.~2, but for any $2n$-sided polygon. 

\begin{figure*}
\includegraphics[width=2in]{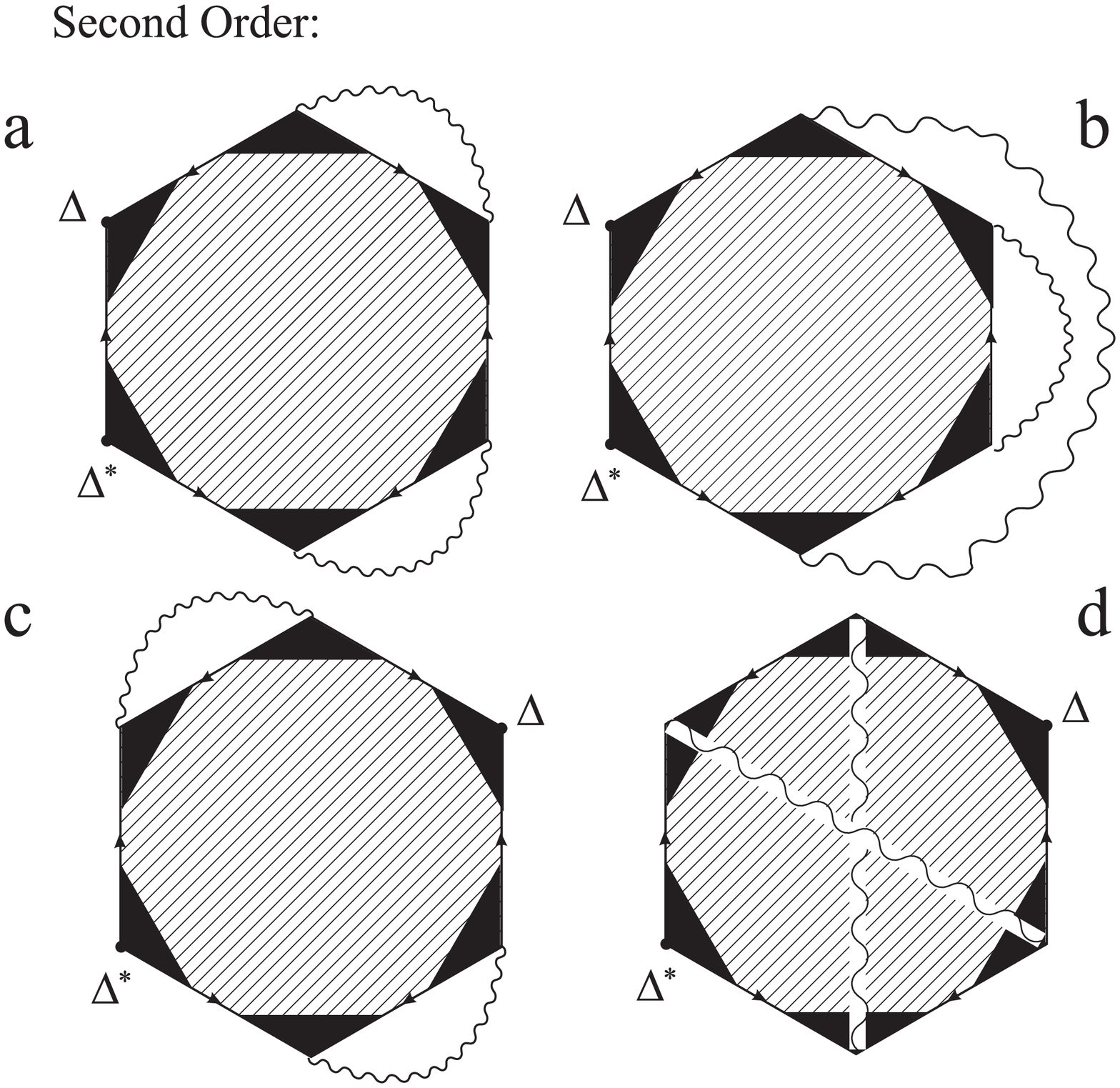}
\caption{\label{fig:fig4} Second order fluctuation corrections to the quadratic part of the Ginzburg-Landau free energy expansion.}
\end{figure*}

\begin{figure*}
\includegraphics[width=4in]{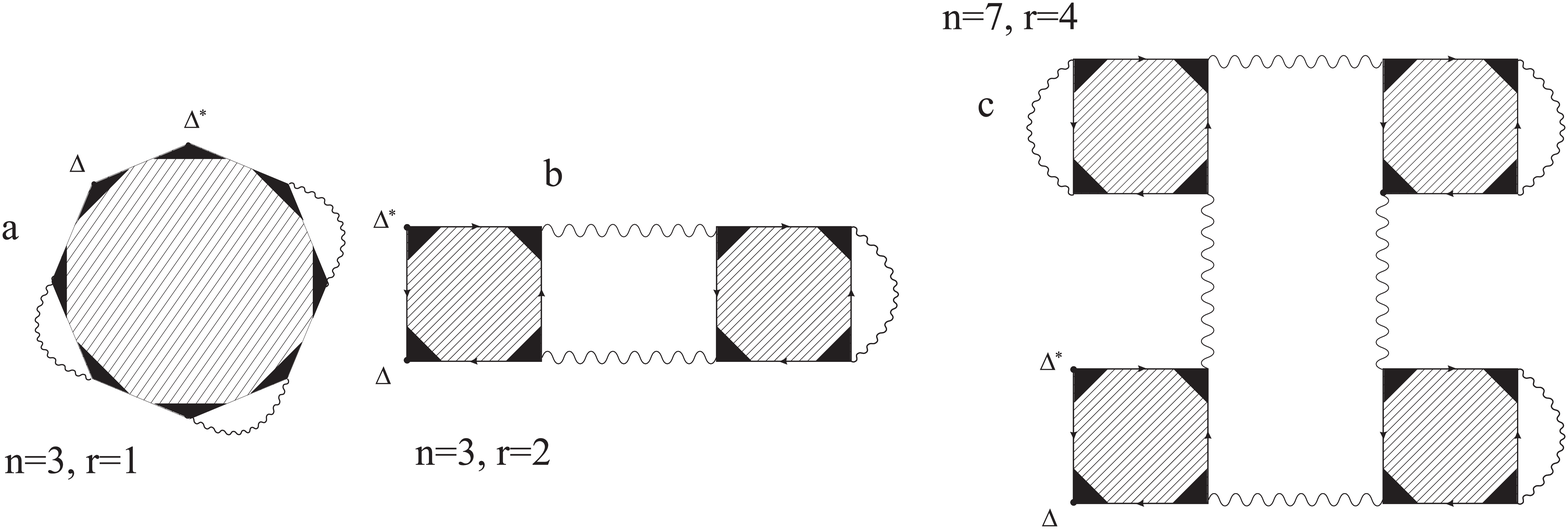}
\caption{\label{fig:fig5} Some higher order diagrams are shown. Graphs ``b'' and ``c'' are examples of reducible diagrams which can be
separated into several irreducible parts by cutting curly lines only.}
\end{figure*}  

Acting by operator (\ref{HB}) on the lowest Landau level eigenfunctions in Eq.(\ref{da}) and calculating
the Matsubara sum, we  derive the following expression:
\begin{eqnarray}
\label{da01}
\delta A_0^{(1)} = -{1 \over 2 \pi L_H^2}\, {N(0) \over \left( 4 \pi T \right)^2 } \psi''\left( {\Omega_H \over 8 \pi T} + {1 \over 2} \right)
\,\left[ T \sum\limits_\Omega {\cal L}\left( \Omega \right) \right].
\end{eqnarray}
The last factor can be evaluated with the logarithmic accuracy as follows:
\begin{eqnarray}
\label{sumL}
\nonumber
T \sum\limits_\Omega {\cal L}\left( \Omega \right) = 
{\Omega_H \over 4 \pi N(0)}
\Biggl[ && \!\!\!\!\!\! \ln{\Omega_H \over T} - \psi\left( {H - H_{c2}(T) \over H_{c2}(0)}{\Omega_H \over T}\right) \\
&& \!\!\!\! - {T \over 2 \Omega_H} { H_{c2}(0) \over H - H_{c2}} \Biggr],
\end{eqnarray}
Let us remember that $N(0)$ is the density of states at the Fermi-line, $\Omega_H = 4 e {\cal D} H = {2 \pi \over \gamma} T_{c0}$, 
where ${\cal D}$ is the diffusion coefficient, $e$ is
the electron charge, $H$ is the external field, $T_{c0}$ is the transition temperature in zero field, and $\gamma \approx 1.78$ is Euler's constant.
$H_{c2}(T)$ stands for the mean field temperature dependence of the upper critical field which is defined by Eqs.(\ref{Ln},\ref{MFeq}) and in
the low-temperature limit has the form:
$$
H_{c2}(T ) \approx H_{c2}(0) \left[ 1 - {2 \gamma^2 \over 3}\, \left({ T \over T_{c0}} \right)^2 \right].
$$

Eqs.(\ref{da01},\ref{sumL})  determine the first order correction to the upper critical field which is defined by the following
equation:
$$
A_0 + \delta A_0^{(1)} = 0.
$$
This is a non-linear algebraic equation, which can be easily solved by numerical means. As one can see in Fig.~3, the first correction
yields  a lower upper critical field  compared to the mean field result.\cite{HW} As expected,\cite{GaL_Reply}  
the perturbative analysis does not lead to an upward curvature in the $H_{c2}(T)$-line.

Let us study the asymptotic behavior of the obtained expressions.
If the following inequality is satisfied $T/T_{c0} \ll \left[ H - H_{c2}(T) \right]  / H_{c2}(0)  \ll 1$, one can obtain the following 
zero-temperature result:
\begin{equation}
\label{1th}
{\delta A_0^{(1)} \over N(0)} = {{\rm Gi}\over 4 \pi}\,  \ln{\left[ H \over H - H_{c2}(0) \right]},
\end{equation}
where we have introduced the dirty limit Ginzburg parameter as ${\rm Gi} = {\left( \varepsilon_{\rm F} \tau \right)^{-1}} \ll 1$.
One can see that the first order correction is only logarithmic.
In the case of small but finite temperatures $\left[ H - H_{c2}(T) \right] / H_{c2}(0)  \ll T/T_{c0} \ll 1$ we have
\begin{equation}
\label{1ht}
{\delta A_0^{(1)} \over N(0)} = {\gamma {\rm Gi}\over \pi^2}  \, {\left[ H \over H - H_{c2}(T) \right]} {T \over T_{c0}}. 
\end{equation}
We see that in  this temperature domain, the effect of the superconducting fluctuations is more pronounced.

\section{Higher order corrections}

In this section, we will study higher order corrections to the mean field result. For the sake of technical simplicity,
we will focus on the zero-temperature case which we will use to estimate the divergence rate of higher order diagrams.
Generalization to non-zero temperatures is straightforward. 

The simplest  way to construct a higher order correction is to study a diagram made of a polygon 
with all possible impurity avergaings and $n$ curly lines connecting $2n$ vertices of the $(2n+2)$-sided polygon.
In Fig.~4, all possible topologically non-equivalent second order diagrams are shown (see also Fig.~5, diagram ``a'').
Note that in the highly disordered limit, these polygons are local quantities which can be written in the form similar to 
(\ref{HB}). The contributions due to such diagrams can be easily estimated. Expressions for these diagrams contain
$n$ integrals over the frequencies running through $n$ fluctuation vertices, which leads
to  a contribution of order of  $\ln^n \left[{H_{c2} \over H - H_{c2}}\right]$. Therefore, we conclude that perturbation 
series due to the single-polygon (irreducible) graphs has the form $\sum_n C_n {\rm Gi}^n \ln^n \left[{H_{c2} \over H - H_{c2}}\right]$, 
where  $n$ is the perturbation series order and $C_n$ are some combinatorial factors. Let us mention that similar diagrams
have been studied by  Kee {\em et al.} \cite{Aleiner} who constructed a non-perturbative resummation technique
to sum up the dominant contributions. In our problem, such a resummation is not possible since, as we shall see,
another type of diagrams delivers the dominant contributions.

Problems start to appear in third order perturbation theory when
one is able to construct a diagram containing several polygons connected by curly lines only (see, {\em e.g.}, Fig.~4; diagram ``b'').
 We will call a diagram reducible if it can be separated into several parts by cutting curly lines only.
 Otherwise, we will call a diagram irreducible. Let us consider a general case of a graph in $n$ order perturbation theory
 which contains $r$ irreducible parts. In this case, simple dimensional analysis of the
 problem yields the conclusion that we have $(n+1-r)$ integrals running through $n$ singular fluctuation propagators.
 Therefore, the contribution due to such a diagram can be estimated as follows:
 $$
{ \delta A^{(n,r)}_0 \over N(0)} \sim {\rm  Gi}^n \left[{H_{c2} \over H - H_{c2}}\right]^{(r-1)} \, \ln^{(n-r)} \left[{H_{c2} \over H - H_{c2}}\right].
 $$
One can easily see that in higher orders the singularity gets enhanced and contains a power low dependence on the closeness
to the transition compared to only logarithmic dependence (\ref{1th}) in first and second order perturbation theory.
At low but finite temperatures  $\left[ H - H_{c2}(T) \right]/ H_{c2}(0)  \ll T/T_{c0} \ll 1$, we obtain the following estimate
for the contribution of an $n$ order graph built up of $r$ irreducible fragments:
 $$
 {\delta A^{(n,r)}_0 \over N(0)} \sim {\rm  Gi}^n  \left( {T  \over T_{c0} } \right)^{(n-r+1)} \left[ { H_{c2}(0)  \over H - H_{c2}(T) } \right]^n.
 $$
It is clear that the more irreducible parts contains a diagram, the more singular contribution it derives. Therefore, mostly divergent
diagrams should contain as many cubic irreducible elements as possible. The corresponding subseries is built up of
chain and ring-like diagrams (see Fig.~4, diagrams ``b'' and ``c''). The contribution from a diagram which contains $r$ Hikami
boxes is found as ($T=0$):
\begin{equation}
\label{dom}
{\delta A^{(2r-1,r)}_0 \over N(0)} \sim  {\rm  Gi}^{2r-1} 
\left[{H_{c2} \over H - H_{c2}}\right]^{r-1}\, \ln^{(r-1)} \left[{H_{c2} \over H - H_{c2}}\right]
\end{equation}
 Such diagrams provide dominant contributions in the expansion.
From Eq.(\ref{dom}), we see that although the singularity gets progressively stronger in the higher orders,  it tends to a finite limit.
Namely one can conclude (by taking the limit $n \to \infty$) that if  
$ \left[ \left(H - H_{c2}\right) / H_{c2} \right]  \ln^{-1} \left[{H_{c2} / \left( H - H_{c2} \right)} \right] \gg {\rm Gi}^2$, each term
in the perturbation series is small. 

Let us mention that keeping any finite number of terms in the perturbation series does not make any qualitative changes
to the shape of the renormalized $H_{c2}(T)$-curve compared to the first correction and to the mean-field results (see Fig.~3).
However, increasing the number of terms in the series leads to a further decrease of the upper critical field.
This gives a very tentative indication that superconducting fluctuations in a two-dimensional system can significantly
change the upper critical field but, probably, should not yield an upward curvature.


\section{Conclusion}

We have considered the effect of superconducting fluctuations on the upper critical field of a two-dimensional disordered
superconductor at low temperatures. We have performed an exact calculation of the first 
perturbation correction and found the corresponding shift of the critical line. We have also estimated 
higher order corrections. The renormalized upper critical field 
was shown to be shifted toward lower fields compared to the classical mean field curve. However, no upward
curvature was found in the framework of any finite-order perturation theory expansion.

We have carefully studied  higher order contributions and found that each consecutive order in the perturbation theory enhances the singularity.
We formulated a general rule of constructing and calculating the dominant contributions in the perturbation series.
 One of our main conclusions is that  the derivation of the Ginzburg-Landau functional 
 in the case of strong fields is hardly possible in the framework of the conventional perturbation theory technique.

Indeed, our results and conclusions are directly applicable for the case of  low temperatures and strong fields only. However, we feel that the issue 
of the perturbation theory applicability and convergence may be essential in the case of weak or even zero field as well. 
Certainly, in this case calculations are technically absolutely different from the case of strong fields, since one has
to evaluate integrals over momenta instead of tracing over the Landau level indices. However, simple estimates show that
reducible graphs bring up more singular contributions compared to the irreducible ones just like in the low-temperature case. 
Moreover, another problem of treating short-wavelength singularities comes out.\cite{pc}

Taking into account these considerations we conclude that the problem of the renormalization of Ginzburg-Landau
coefficients by superconducting fluctuations and, therefore, the renormalization of the transition point itself, are essentially non-perturbative
problems in a two-dimensional case. This makes the Ginzburg-Landau approach ill-justified. By saying this, we certainly do not
challenge the commonly accepted and well-tested phenomenological form of the Ginzburg-Landau functional. However, it is possible
that the coefficients in the expansion are quite different from the conventional mean-field prediction due to the non-perturbative
renormalization by superconducting fluctuations. 

\begin{acknowledgements}
This work was supported by the US-ONR, the NSF, the LPS, and DARPA. V.G. wishes to thank 
A. A. Varlamov  and D. J. Priour, Jr. for  useful discussions. V.G. is especially grateful 
to Alex Kamenev for pointing out a sign error in the original version of 
the manuscript.
\end{acknowledgements}


\begin{thebibliography}{}

\bibitem{exp} See {\it e.g.:} S. Okuma, J. Phys. Soc. Jpn. {\bf 52}, 3269 (1983);
A. F. Hebrad and M. A. Paalanen, Phys. Rev. B {\bf 30}, 4063
(1984).

\bibitem{HW} E. Helfand and N. R. Werthamer, Phys. Rev. {\bf 147},  288 (1966).

\bibitem{GD} A.A. Golubov and V.V. Dorin, J. Low. Temp. Phys. {\bf 78}, 378 (1990).

\bibitem{Smith} R. A. Smith, B. S. Handy, and V. Ambegaokar Phys. Rev. B
{\bf 61}, 6352 (2000).

\bibitem{OF} Y. Oreg and A.M. Finkel'stein, Phys. Rev. Lett. {\bf 83}, 191 (1999).

\bibitem{ZS} F. Zhou and B. Spivak, Phys. Rev. Lett. {\bf 74}, 2800 (1995).

\bibitem{GaL} V. M. Galitski and A. I. Larkin, Phys. Rev. Lett. {\bf 87},  087001 (2001).

\bibitem{Ikeda_Com}R. Ikeda, Phys.Rev.Lett. {\bf 89}, 109703(2002).

\bibitem{GaL_Reply} V. M. Galitski and A. I. Larkin, Phys. Rev. Lett {\bf 89}, 109704 (2002)


\bibitem{MW} N. D. Mermin and H. Wagner, Phys. Rev. Lett. {\bf 17}, 1133 (1966); P. C. Hohenberg, Phys. Rev.
{\bf 158} 383 (1967).

\bibitem{LV} A. I. Larkin and A. A. Varlamov in ``Handbook on Superconductivity: Conventional and Unconventional Superconductors'' edited by
     K.-H.Bennemann and J.B. Ketterson, Springer, 2002; [cond-mat/0109177].

\bibitem{deG}C. Caroli, M. Cyrot and P. G. de Gennes, Solid State Commun. {\bf 4}, 17 (1966); see also:
 K. Maki, Phys. Rev. {\bf 148}, 362 (1966).

\bibitem{Aleiner} Hae-Young Kee, I.L. Aleiner, and B.L. Altshuler, Phys. Rev. B {\bf 58}, 5757-5776 (1998).

\bibitem{pc} A. A. Varlamov (private communication).

\end{thebibliography}
\end{document}